\documentclass[british,epjST]{svjour}
\usepackage[T1]{fontenc}
\usepackage[latin9]{inputenc}
\usepackage{color}
\usepackage{amsmath}
\usepackage{amssymb}
\usepackage{graphicx}
\usepackage{babel}
\begin{document}

\title{Decoherence effects in the dynamics of interacting ultracold bosons
in disordered lattices}

\author{Beno\^{i}t Vermersch \and Jean~Claude Garreau\thanks{http://www.phlam.univ-lille1.fr/atfr/cq}}
\institute{Laboratoire
de Physique des Lasers, Atomes et Mol\'ecules, Universit\'e Lille 1 Sciences
et Technologies, CNRS; F-59655 Villeneuve d'Ascq Cedex, France}
\abstract{
We study the interplay between disorder, interactions and decoherence
induced by spontaneous emission process. Interactions are included
in the Anderson model via a mean-field approximation, and a simple
model for spontaneous emission is introduced.
Numerical simulations allow us to study the effects of decoherence
on different dynamical regimes. Physical pictures for the mechanisms
at play are discussed and provide simple interpretations. Finally,
we discuss the validity of scaling laws on the initial state width.
}
\maketitle
\section{Introduction}

The Anderson model \cite{Anderson:LocAnderson:PR58}, is a simple,
tractable, model describing disorder effects on electrons in a crystal.
The model predicts that in disordered crystals the eigenfunctions
may become spatially localised %
\footnote{In 1D all eigenfunctions are localised, in 3D an energy {}``mobility
edge'' separates localised from delocalised ones.%
}, in sharp contrast with the Bloch functions of a perfect crystal,
a phenomenon called \emph{Anderson localisation}. However, due to
its very simplicity, the model (in its original form) overlooks potentially
important effects: It is a single particle, zero temperature model.

Recently, analogs of the Anderson model have been realized with cold
or ultracold atoms, which can be made much closer to the Anderson
model than a crystal. A direct translation of the Anderson model can
be obtained by placing ultracold atoms in a far-detuned laser speckle,
which realizes a random mechanical potential affecting the center-of-mass
of the atom, and allowed observation of the localisation in one dimension
(1D) with bosons \cite{Billy:AndersonBEC1D:N08,Inguscio:AubryAndreInteractions:NP10}.
The 3D localisation was also studied with both fermions \cite{Kondov:ThreeDimensionalAnderson:S11}
and bosons \cite{Jendrzejewski:AndersonLoc3D:NP12}. The quasiperiodic
kicked rotor can also realize an equivalent of both 1D \cite{Moore:AtomOpticsRealizationQKR:PRL95,Lignier:Reversibility:PRL05,Chabe:PetitPic:PRL06}
and 3D systems \cite{Casati:IncommFreqsQKR:PRL89,Chabe:Anderson:PRL08};
the latter allowing a detailed experimental study of the Anderson
\emph{transition}, including measurements of the critical wavefunction
\cite{Lemarie:CriticalStateAndersonTransition:PRL10} and of its critical
exponent \cite{Chabe:Anderson:PRL08,Lopez:ExperimentalTestOfUniversality:PRL12,Lemarie:AndersonLong:PRA09}.
Ultracold atom systems allow a large control of decoherence, which
is essentially due to spontaneous emission, and of atom-atom interactions,
for which an experimental {}``knob'' is provided by the Feshbach
resonances \cite{Chin:FeshbachResonances:RMP10}, making them an ideal
tool for studies of the interplay of quantum effects, disorder, interactions,
and decoherence.

We have recently studied numerically and theoretically the influence
of interactions in a bosonic 1D Anderson system \cite{Vermersch:AndersonInt:PRE12},
which is a very active research domain \cite{Shepelyansky:DisorderNonlin:PRL08,Flach:DisorderNonlineChaos:EPL10}.
The sensitivity of the system to the initial state was shown to be
characterised by a scaling law \emph{depending on the initial state}.
In the present paper we introduce the effect of decoherence due to
spontaneous emission in the problem, and show that Anderson localisation
(AL) is promptly destroyed by it, but the resulting dynamics still
respects the scaling law mentioned above. This confirms the pertinence
of the scaling strategy, and indicates that it is a relevant approach for studying disordered interacting systems.

\section{The model\label{sec:model}}

We essentially use in the present paper the same model as in \cite{Vermersch:AndersonInt:PRE12},
except for the introduction of decoherence effects as discussed below.
Briefly, we start with a gas of interacting bosons evolving in the
sinusoidal potential formed by a 1D standing wave $V(x)=-V_{0}\cos^{2}(k_{L}x)$.
In the mean field approximation, the ultracold gas can be represented
by a single wave-function $\psi_{\epsilon}(x)$ obeying the Gross-Pitaesvkii
equation (GPE): 
\[
i\hbar\dot{\psi}_{\epsilon}(x)=\left(\frac{p^{2}}{2m}+V(x)+g|\psi_{\epsilon}(x)|^{2}\right)\psi_{\epsilon}(x)
\]
Projecting the solution on a basis of localised Wannier functions
$w_{n}(x)$, $\psi_{\varepsilon}(x)=\sum_{n}c_{n}(t)w_{n}(x)$ leads
to the well-known \emph{tight-binding} description: 
\begin{equation}
i\hbar\dot{c}_{n}=v_{n}c_{n}-t_{n-1}c_{n-1}-t_{n+1}c_{n+1}+g\left|c_{n}\right|^{2}c_{n},\label{eq:DNSE}
\end{equation}
where $g$ represents the interaction strength which is proportionnal
to the s-wave scattering length, and we kept only nearest-neighbors
hopping terms, i.e. $t_{n+m}=0$ if $|m|>1$. According to standard
conventions, the hopping term is symmetric $t_{n+1}=t_{n-1}$$=T$
and we measure energies in units of $T$ and time in units of $\hbar/T$.
Following the Anderson {}``recipe'' \cite{Anderson:LocAnderson:PR58}
we introduce {}``diagonal'' disorder by randomizing the onsite energies
$v_{n}$ in an interval $[-W/2,W/2]$; each choice of an ensemble
$\{v_{n}\}$ producing a {}``realization'' of the disorder, so,
finally: 
\begin{equation}
i\dot{c}_{n}=v_{n}c_{n}-c_{n-1}-c_{n+1}+g\left|c_{n}\right|^{2}c_{n}.\label{eq:DANSE}
\end{equation}
As we use mean field theory throughout this work, we shall use the
terms {}``interaction'' and {}``nonlinearity'' interchangeably.

Spontaneous emission (SE) is inevitably -- although controllable --
present in an optical potential. For a laser-atom detuning $\Delta=\omega_{L}-\omega_{0}\gg\Gamma_{0}$
(where $\omega_{L}$ is the laser frequency and $\omega_{0}$ the
atomic transition frequency and $\Gamma_{0}$ the natural width of
the transition) the amplitude $V_{0}$ of the potential is $V_{0}=\hbar\Omega^{2}\text{/8\ensuremath{\Delta}}$,
where $\Omega$ is the resonance Rabi frequency. The SE ratio is then
given by 
\[
\Gamma_{\mathrm{se}}=\frac{\Gamma_{0}}{4}\frac{\Omega^{2}}{\Delta^{2}+\Omega^{2}/2+\Gamma_{0}^{2}/4}\sim\frac{\Gamma_{0}\Omega^{2}}{4\Delta^{2}}
\]
where the approximate expression corresponds to the large-detuning
limit $|\Delta|\gg\Gamma_{0},\Omega$. For common parameters used
in experimental setups, $\Gamma_{\mathrm{se}}\lesssim10$s$^{-1}$,
$T/\hbar\sim1000$s$^{-1}$ so that in our rescaled units, the spontaneous
emission rate $\gamma$ in reduced units is typically $\gamma\lesssim10^{-2}$.

We model the effect of spontaneous emission by a Monte Carlo procedure
simulating spontaneous emission events according to the probability
$\gamma$. If a SE event happens, as we are considering a one-dimensional
lattice, we simply translate the momentum $p$ \emph{of the condensate}
(i.e. the order parameter Fourier transform $\psi(p)$) by a vector
$\hbar k_{L}\cos\theta$, or, in reduced units $\pi\cos\theta$,
with $\theta$ being a randomly picked angle (in the interval
$[0,2\pi]$) representing the direction of the emitted photon with
respect to the direction of the laser beam. Between two spontaneous
emission events, the evolution of $c_{n}$ is calculated from Eq.~\eqref{eq:DANSE}.
This way of modeling SE has been often used for \emph{individual atoms}
in optical lattices \cite{Cohen:LocDynTheo:PRA91,Lemarie:AndersonLong:PRA09,Lepers:SupprSpontEm:PRA10}
and compares very well with experimental signals. It can however be
asked if it can also be applied to the \emph{condensate} wavefunction
in the mean-field approximation. This can be justified, at least for
low SE rates, by the following argument. In a condensate, a photon
can be absorbed in two ways: either an individual atom aborbs it and
goes to an excited state, or it is absorbed collectivelly generating
a {}``delocalised'' excitation. In the former case, the individual
atom is not anymore on the condensate and does not contribute to the
mean-field dynamics described by the Gross-Pitaevskii equation; in
the latter case there is no changing in the condensate population
and it is reasonable to consider that it is the \emph{collective}
mean-field wavefunction that {}``recoils'' to compensate the photon
momentum %
\footnote{This process is somewhat analogous to what happens in the Mossbaeur
effect.%
}. Our model of the spontaneous emission corresponds to a collective
recoil of the condensate wavefunction (and to a progressive increasing
of the non-condensate fraction which is not taken into account) an
approximation which is valid as long as this fraction stays small,
that is, for low SE rates compared to the duration of the experiment.

As in ref.~\cite{Vermersch:AndersonInt:PRE12}, we work with a system
of fixed size of $L$ sites, with absorbing boundaries to prevent
wavepacket reflections. This is done by setting a slowly varying imaginary
potential at the edges of the box. The wavepacket is thus depleted
if it is (or becomes) large enough to touch the borders of the box,
and the survival probability $p$ is the integral of the wavepacket
still present inside the box at time $t$:
\[
p(g,t)=\sum_{n=-L/2}^{L/2}\left|c_{n}(t)\right|^{2}.
\]
A diffusive dynamics is thus characterised by a continuous
decrease of $p(g,t)$ with $t$. In the present work typically $L=101$
sites. 

In presence of interactions, the system is nonlinear and thus sensitive
to the initial state. In order to characterise the resulting dependence
of the dynamics on the size of the initial state, we study a particular
family of initial states of square shape and width $L_{0}$:
\begin{equation}
c_{n}(0)=\begin{cases}
L_{0}^{-1/2} & \exp\left[i\theta_{n}\right],\ |n|\le\left(L_{0}-1\right)/2\\
0 & \mathrm{otherwise.}
\end{cases}\label{eq:InitialState}
\end{equation}
 with \emph{random} phases $\theta_{n}$ (for a discussion of the
consequences of this choice, see~\cite{Vermersch:AndersonInt:PRE12}).

\section{Decoherence effects in disordered, interacting systems}

In a tight-binding approach, one associates eigenenergies and eigenstates
to lattice sites. In a periodic lattice, each eigenenergy matches
a degenerate correspondant in a neighbor site, which allows tunneling
and diffusion. In presence of diagonal disorder, this degeneracy is
removed, and only \emph{virtual} tunneling is allowed, which leads
to the exponentially localised Anderson eigenstates. However, spontaneous
emission can couple Anderson eigenstates even if they are not degenrate,
and thus induces diffusive motion in a disordered lattice.

Figure~\ref{fig:pvsgamma} shows the survival probability $p$ at
time $t=10^{5}$ as a function of the spontaneous emission rate $\gamma$
for five values of the disorder $W$, in absence of interactions ($g=0$).
Data were averaged over a total of 500 realizations of the disorder
$\{v_{n}\}$, of the inital phase distribution $\{\theta_{n}\}$\textcolor{red}{{}
}and of SE emission times. At time $t$, there have been on the average
$\gamma t$ spontaneous emission events. The number of spontaneous
emission events necessary for destroying AL depends on the disorder
$W$. For $W=2$ (green triangles), 10 spontaneous emissions are enough
to induce complete diffusion of the wavepacket whereas for $W=8$
(yellow inverted triangles) the wavepacket escapes from the box after
$10^{4}$ events ($\gamma=10^{-1}$).

One can interpret these results in two equivalent ways. A {}``static''
point of view consists on noting that the initial state {[}Eq.~(\ref{eq:InitialState}){]},
due to the choice of random phases, projects over \emph{all} Anderson
eigenvalues. These eigenstates have different localisation lengths,
with a maximum at the center of the band given by $\ell_{0}\sim96W^{-2}$
\cite{Luck:SystDesord:92}, and zero at the borders of the band. If
the disorder is small, a typical Anderson eigenstate has width comparable
or larger than the box width $L=101$, it is thus absorbed in a short
time. The initial survival probability then corresponds to the fraction
of Anderson eigenstates with localisation length much smaller that
$L$. However, when a spontaneous emission event happens, it \emph{redistributes}
the quantum phases, and projects again a part of the wavepacket on
eigenstates that are either very large or centered in positions close
to the border of the box, which are then absorbed, and this repeated
process progressively leads to the destruction of the AL. A dynamic
point of view considers that the initial state evolves during a characteristic
localisation time $t_{\mathrm{loc}}\sim W^{-2}$ until it localises,
with a typical width $\left\langle x^{2}\right\rangle \sim W^{-4}$.
The phase redistribution due to a spontaneous emission event breaks
the quantum interference responsible for the localisation, diffusion
starts again for a time $t_{\mathrm{loc}}$, and then stops until
the next SE event, and so on. For low SE ratios $\gamma t_{\mathrm{loc}}\le1$, if $D$ is the disordered-averaged \emph{classical} diffusion coefficient,
then one can estimate the diffusion coefficient induced by SE as $D_{\mathrm{SE}}=Dt_{\mathrm{loc}}\gamma$.

\begin{center}
\begin{figure}
\centering{}
\includegraphics[width=0.5\columnwidth]{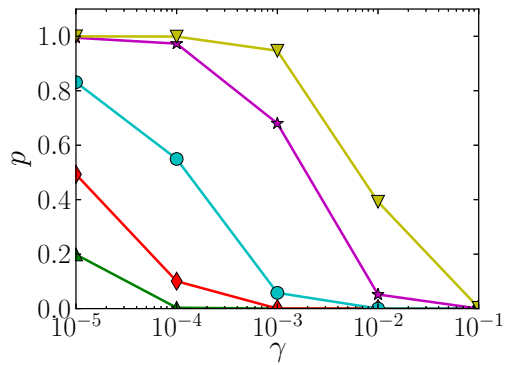}
\caption{\label{fig:pvsgamma}(Color online) Survival probability $p$ as a
function of the spontaneous emission rate $\gamma$ at $t=10^{5}$
in the absence of interactions ($g=0$). Values of the disorder $W$
are 2 (green triangles), 3 (red diamonds), 4 (cyan
circles) , 6 (magenta stars) and 8 (yellow inverted triangles). The
width of the initial state is $L_{0}=21$.}
\end{figure}

\par\end{center}

Figure~\ref{fig:phivsn} presents spatial probability distributions
$\left|c_{n}\right|^{2}$ for $t=10^{5}$, $W=4$, $L_{0}=3$ for
various values of the spontaneous emission rate $\gamma$. For $\gamma=0$
(blue solid line) this distribution is exponentially localised $|c_n|^2\propto\exp\left(-2\left|n\right|/\ell\right)$,
where $\ell$ is the localisation length, comparable to $\ell_{0}(W)$,
the maximum localisation length of the Anderson eigenstates. For $\gamma=10^{-5}$
(green dotted line), the system is only weakly affected by spontaneous
emission (as just one event happens in average during the time evolution),
a small diffusion is observed in the tails of the distribution. For
$\gamma=10^{-4}$ (red dashed line) and $\gamma=10^{-3}$ (dot-dashed
cyan line), the distribution has taken a Gaussian shape, indicating
that the decoherence-induced diffusion has become dominant. Alternatively,
this process can be interpreted {}``statically'' as a phase reshuffling
provoked by spontaneous emission, which excites eigenstates
centered everywhere in the box $L$, so that the localisation length
is not anymore a relevant length scale in the problem. 

\begin{center}
\begin{figure}
\centering{}
\includegraphics[width=0.5\columnwidth]{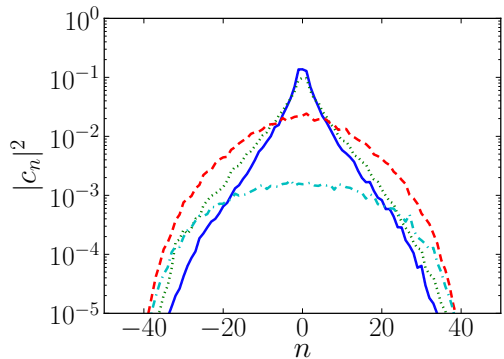}
\caption{\label{fig:phivsn}(Color online) Density distribution $|c_{n}|^{2}(t=10^{5})$
\emph{vs} site index $n$, for $g=0$, $W=4$, $L_{0}=3$. Values
of $\gamma$ are: $0$ (blue solid line), $10^{-5}$ (green dotted
line), $10^{-4}$ (red dashed line) and $10^{-3}$ (dot-dashed cyan
line). Anderson localisation, characterised by the exponential (triangular
in logarithmic scale) shape is progressively destroyed when decoherence
increases.}
\end{figure}

\par\end{center}

\begin{figure}
\begin{centering}
\includegraphics[width=0.5\columnwidth]{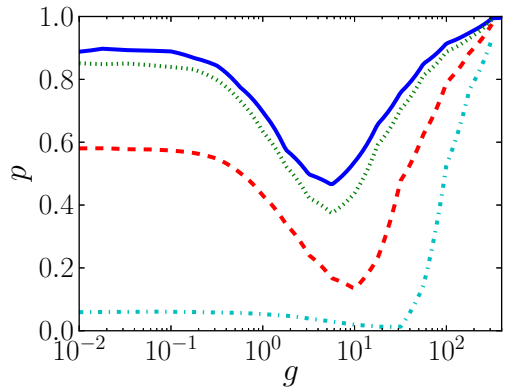}
\par\end{centering}

\caption{\label{fig:pvsg}(Color online) Survival probability $p$ as a function
of the interacting strength $g$ for $t=10^{5}$. Values of $\gamma$:
0 (blue solid line), $10^{-5}$ (green dotted line), $10^{-4}$ (red
dashed line) and $\gamma=10^{-3}$ (dot-dashed cyan line). Three dynamical
regimes are clearly visible: Localized (low $g$), chaotic (intermediate
values of $g$), and self-trapped (high $g$). Other parameters are
$W=4$ and $L_{0}=3$.}

\end{figure}

Consider now the repulsive interacting case ($g>0$). It is useful, if not completely rigorous, to interpret the nonlinear term in Eq.~\eqref{eq:DANSE},
as a \emph{dynamical perturbation} $v^{\mathrm{NL}}\equiv g|c_{n}|^{2}$
of the on-site energy $v_{n}$, depending on the population of the
site. For $v^{NL}\ll W$ -- which implies a low-density, spatially
extended, wavepacket -- Anderson localisation is expected to survive
for a very long time: Anderson eigenstates whose localisation length
$\ell$ is inferior to the size of the box remain in the system and
give non-zero survival probability. For $v^{\mathrm{NL}}\sim W$,
if the $c_{n}(t)$ vary rapidly, the dynamical correction can bring
temporarily neighbour levels close to degeneracy, restablishing transport.
This mechanism is most efficient in the so-called \emph{chaotic regime},
where the $c_{n}(t)$ display a stochastic character. If $v^{\mathrm{NL}}\gg W$,
neighbour sites are decoupled from each other even for small population
differences, which leads to a different, nonlinear, type of localisation
called \emph{self-trapping}, favored for narrow (small $L_{0}$) initial
states. Figure~\ref{fig:pvsg} displays the survival probability
$p(t=10^{5})$ as a function of the interacting strength $g$, for
$W=4$ and four values of the decoherence level $\gamma$ from 0 to
$10^{-3}.$ The three dynamical regimes are clearly visible, a plateau
for small values of $g$ due to surviving AL, a dip created by the
chaotic diffusion for intermediate values of $g$, and a reduction
of diffusion due to self-trapping for high values of $g$, leading
to a complete freezing for $g\approx300$. Clearly, the three dynamical
regimes are not affected in the same way by decoherence: Anderson
localisation, which relies on delicate phase relations is more sensitive
to decoherence. Comparison of the curves for $\gamma=0$ (solid blue)
and $\gamma=10^{-5}$ (dotted green) shows that even \emph{one} SE
event in the average has an observable effect on AL. In the chaotic
regime, the diffusion due to chaotic motion and diffusion due to decoherence
tend to add to each other, but a strong diffusion due to decoherence
decreases the site populations, and thus the nonlinear term $g\sum_{n}|c_{n}|^{2}$.
This effect of dilution due to decoherence explains the fact that
the frontier between the various regimes is shifted to higher values
of $g$ when $\gamma$ increases. Self-trapping, not relying on quantum
inteference, but only on populations $|c_{n}|^{2}$ (which are not
directly affected by SE) tends to be insensitive to decoherence: for
$g\approx320$ all curves converge to unitary survival probability.

\begin{center}
\begin{figure}
\includegraphics[width=0.45\columnwidth]{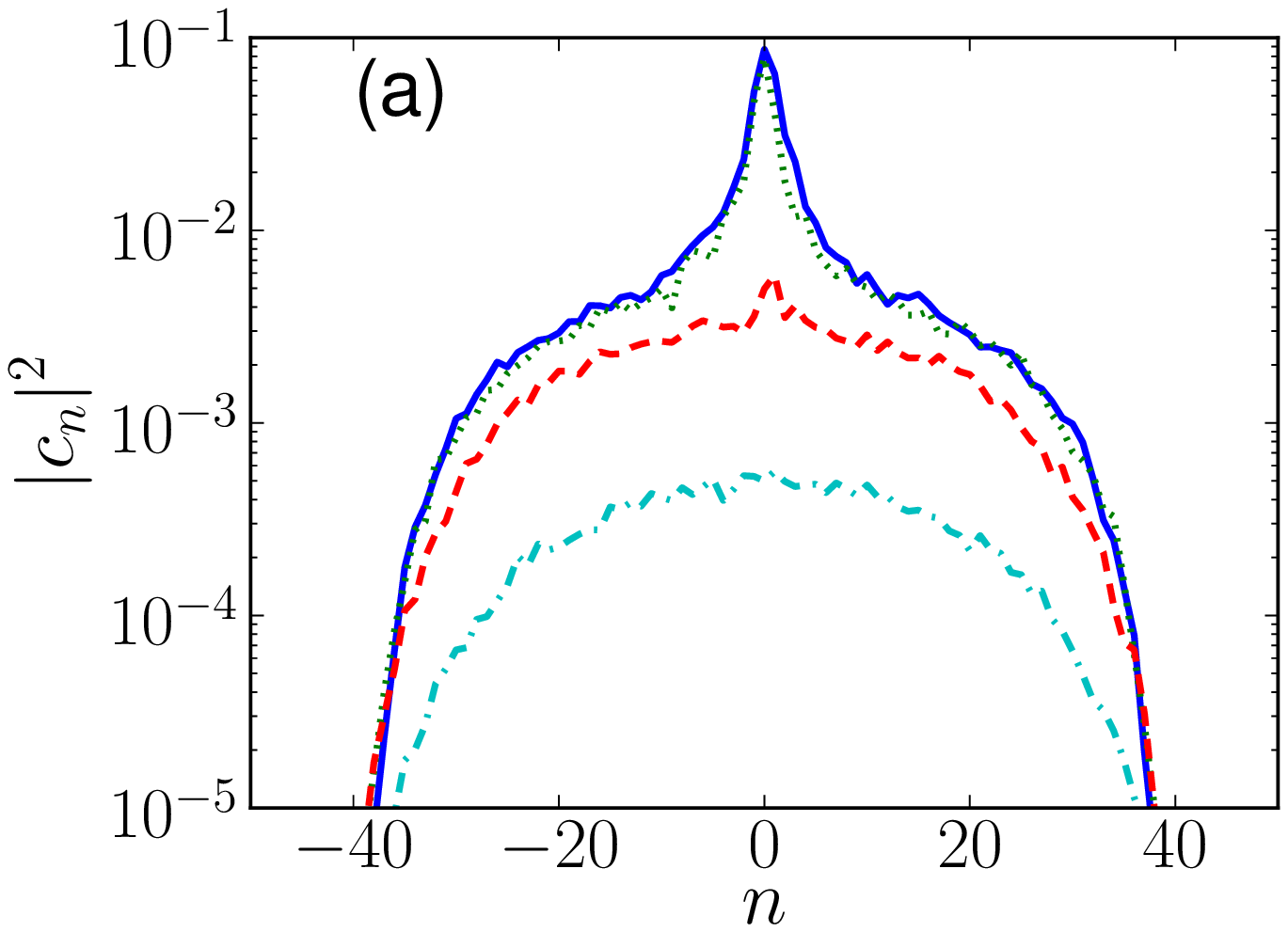}\qquad
\includegraphics[width=0.45\columnwidth]{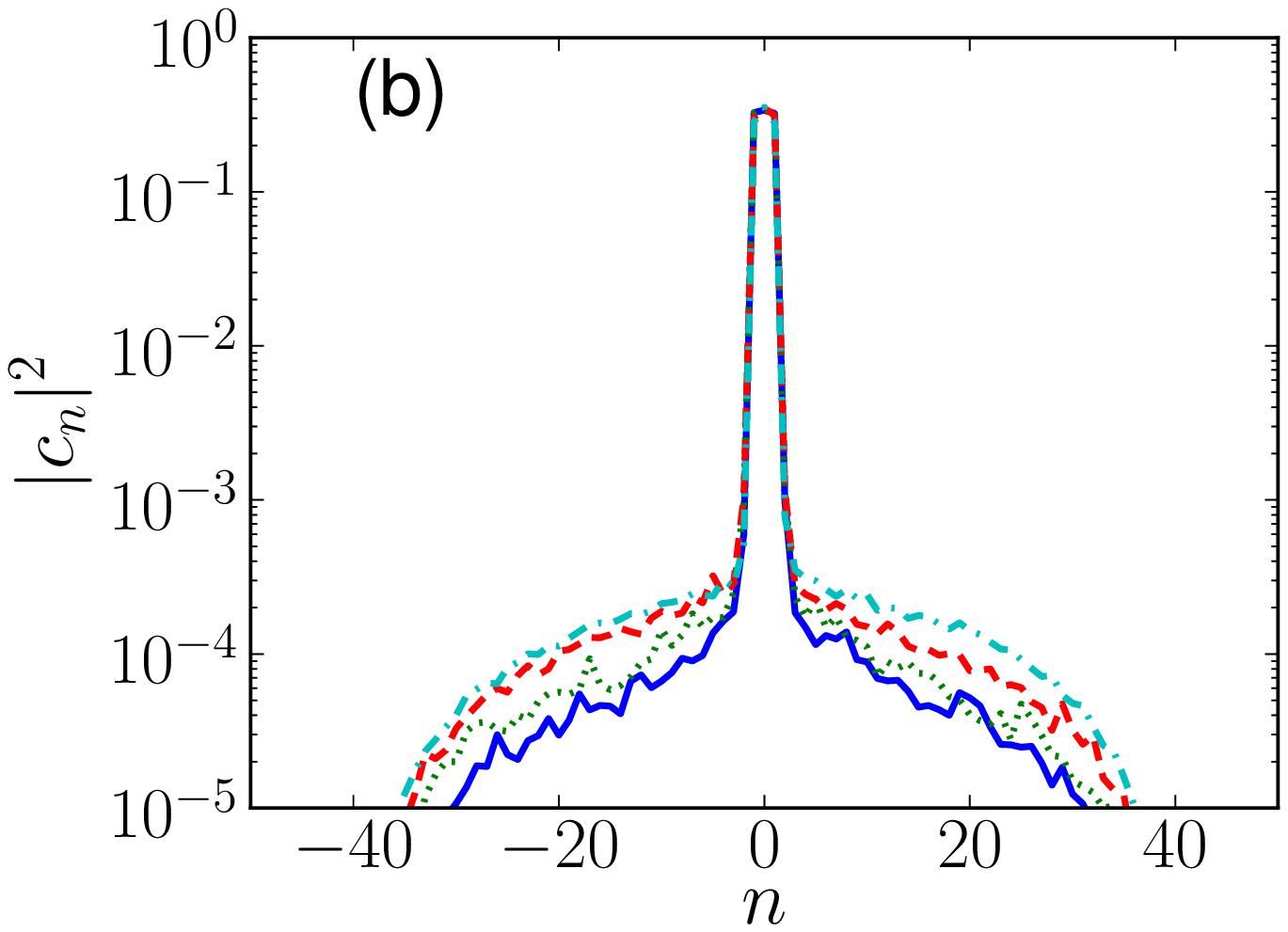}

{}\caption{\label{fig:phivsng}(Color online) Spatial distributions at $t=10^{5}$
for $g=10$(a) and $g=320$(b), and $\gamma=0$ (solid blue line),
$\gamma=10^{-5}$ (green dotted), $\gamma=10^{-4}$ (red dashed) and
$\gamma=10^{-3}$ (dot-dashed cyan). For low nonlinearities and low
decoherence levels, Anderson localisation survives for long times,
but is finally destroyed by decoherence. The self-trapping regime
is almost insensitive to decoherence. Other parameters are $W=4$,
$L_{0}=3$.}
\end{figure}

\end{center}

Figure~\ref{fig:phivsng} shows spatial distributions at $t=10^{5}$
for $g=10$ (a) and $g=320$ (b). In the chaotic regime (a) and for
low decoherence rate (solid blue, dotted green lines), the wavepacket
has its exponential shape preserved in the center of the box but non-exponential
tails appear. This is due to the fact that interactions depopulate
the center of the box (where populations and thus nonlinearity is
maximum), but it preserves the exponential shape \cite{Vermersch:AndersonInt:PRE12}.
The question whether Anderson localisation is destroyed in this case
is tricky: The wavepacket spreads along the box but preserves its
typical exponential-shape. Adding decoherence to the system gives
a far more solid answer to this question. One sees in Fig.~\ref{fig:phivsng}
that the exponential tail peak is almost completely destroyed after 10
spontaneous emissions on the average (dashed red and cyan dot-dashed
lines), which strongly suggested that it is indeed due to interference
effects and can thus be attributed to a survival of AL. On the other
hand, as expected, the self-trapping regime Fig.~\ref{fig:phivsng}b
is very robust against decoherence. 

\section{Scaling laws}

\begin{figure}
\begin{centering}
\includegraphics[width=0.45\columnwidth]{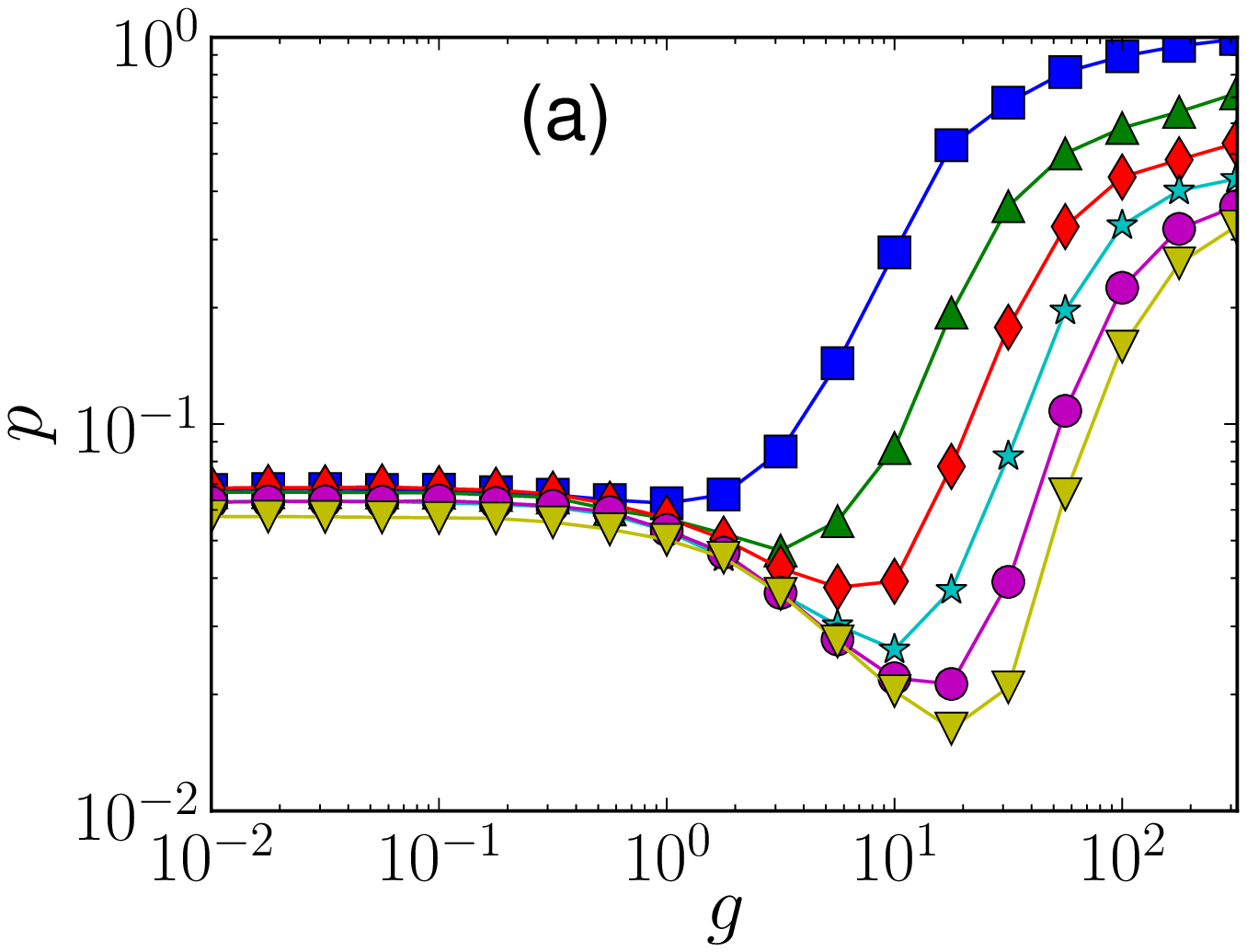}\qquad
\includegraphics[width=0.45\columnwidth]{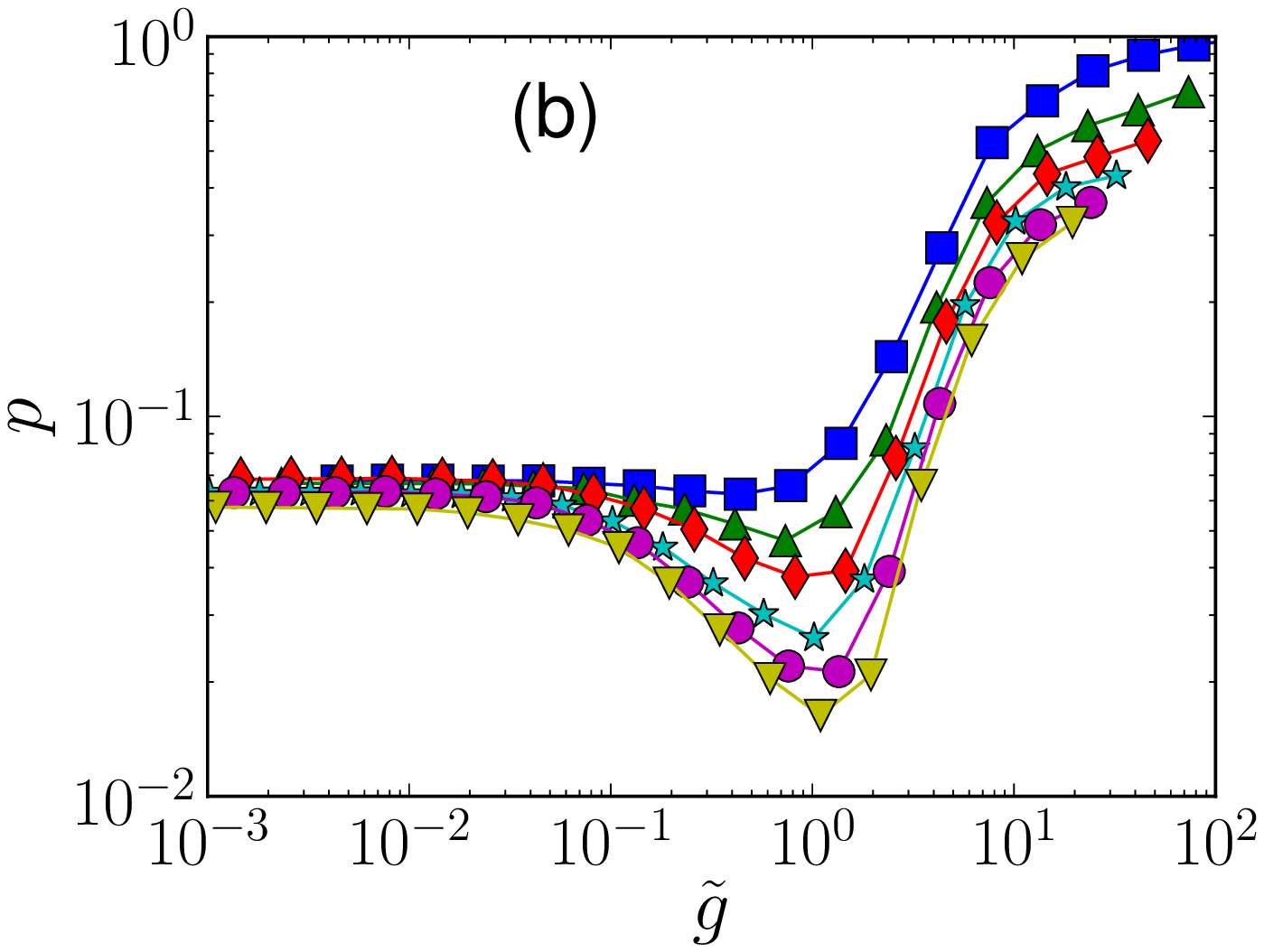}
\end{centering}

\caption{\label{fig:pvsgL0}(Color online) Survival probability as a function
of the nonlinearity for $W=1$, $\gamma=10^{-5}$. (a) No scaling.
(b) The scaling in $\tilde{g}=gL_{0}^{-s}$ ($s=0.76$) allows a clear
identification of regions corresponding to localised ($\tilde{g}<0.1$),
chaotic ($0.1<\tilde{g}<10$) and self-trapped ($\tilde{g}>10$) behaviour.
Values of $L_{0}$ : 3 (blue squares ), 7 (green triangles), 13 (red
diamonds), 21 (cyan stars), 31 (magenta circles), 41 (yellow inverted
triangles).}

\end{figure}

Sensitivity to the initial conditions is a common feature of nonlinear
systems. The nonlinear correction $v^{\mathrm{NL}}$ obviously depends
on the local density and is therefore highly sensitive to the initial
width of the packet: the linear regime is favored for high values
of $L_{0}$ whereas the chaotic regime and self-trapping regime are
expected for spatially concentrated wavepackets. In ref.~\cite{Vermersch:AndersonInt:PRE12},
we showed the existence of scaling laws \emph{on the width $L_{0}$
of the initial wavepacket}, which in particular allowed us to classify
the three regimes as a function of a \emph{scaled interaction strength}
$\tilde{g}=gL_{0}^{-s}$ with $s=0.76\pm0.08$. Figure~\ref{fig:pvsgL0}a,
compares the survival function $p(t=10^{5})$ for different values
of the initial state width $L_{0}$; one clearly sees that the crossovers
between the quasilocalised, chaotic and self-trapped regimes indeed
strongly depend on $L_{0}$. Figure~\ref{fig:pvsgL0}b displays the
same curves plotted in terms of $\tilde{g}$, showing that the crossovers
become independent of the initial state, even in the presence of spontaneous
emission. As opposed to interactions, decoherence effects are not sensitive
to the initial state so it is not surprising that the renormalized nonlinearity
remains a {}``good'' parameter.

\begin{figure}
\includegraphics[width=0.3\columnwidth]{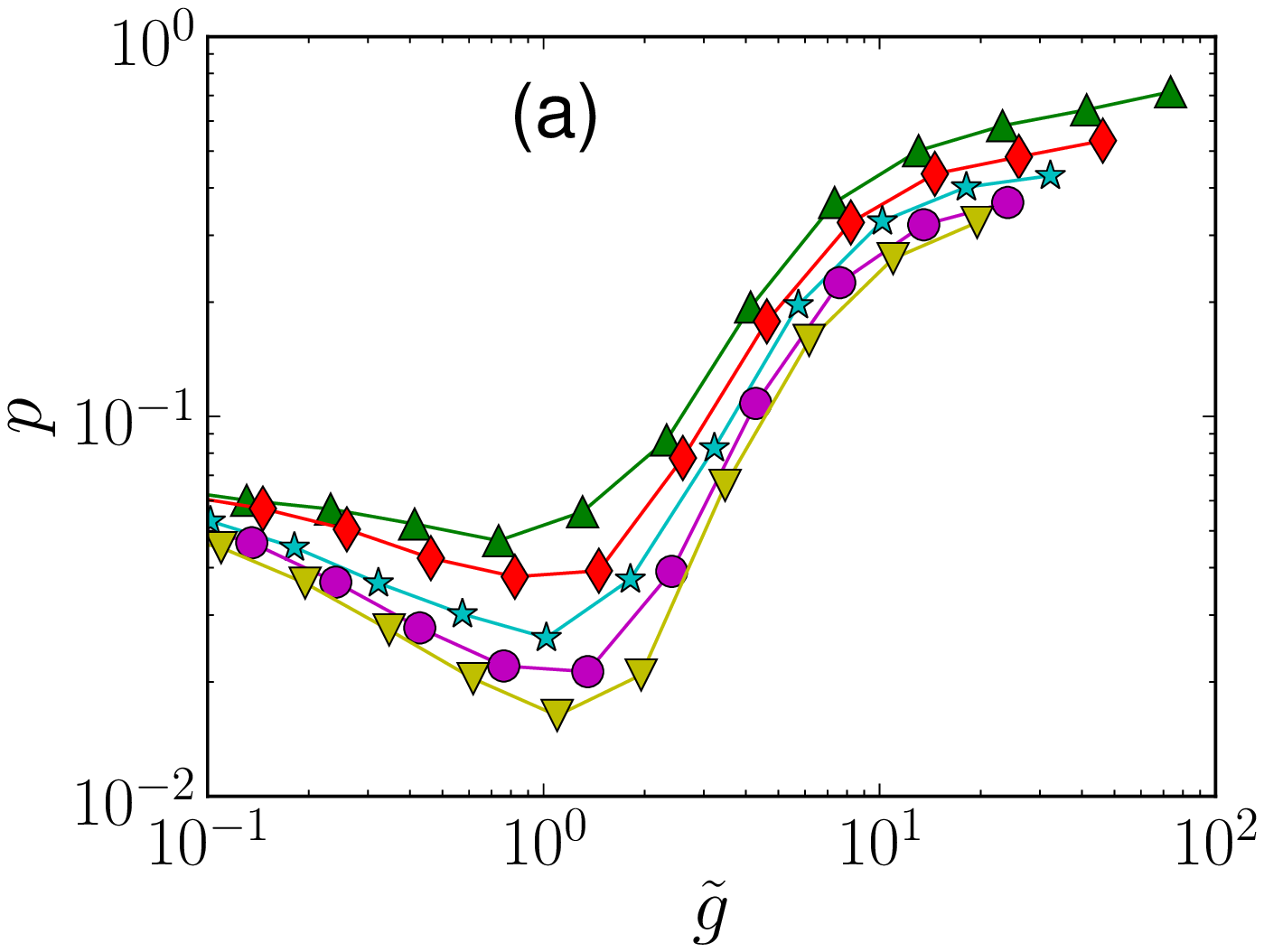}\quad
\includegraphics[width=0.3\columnwidth]{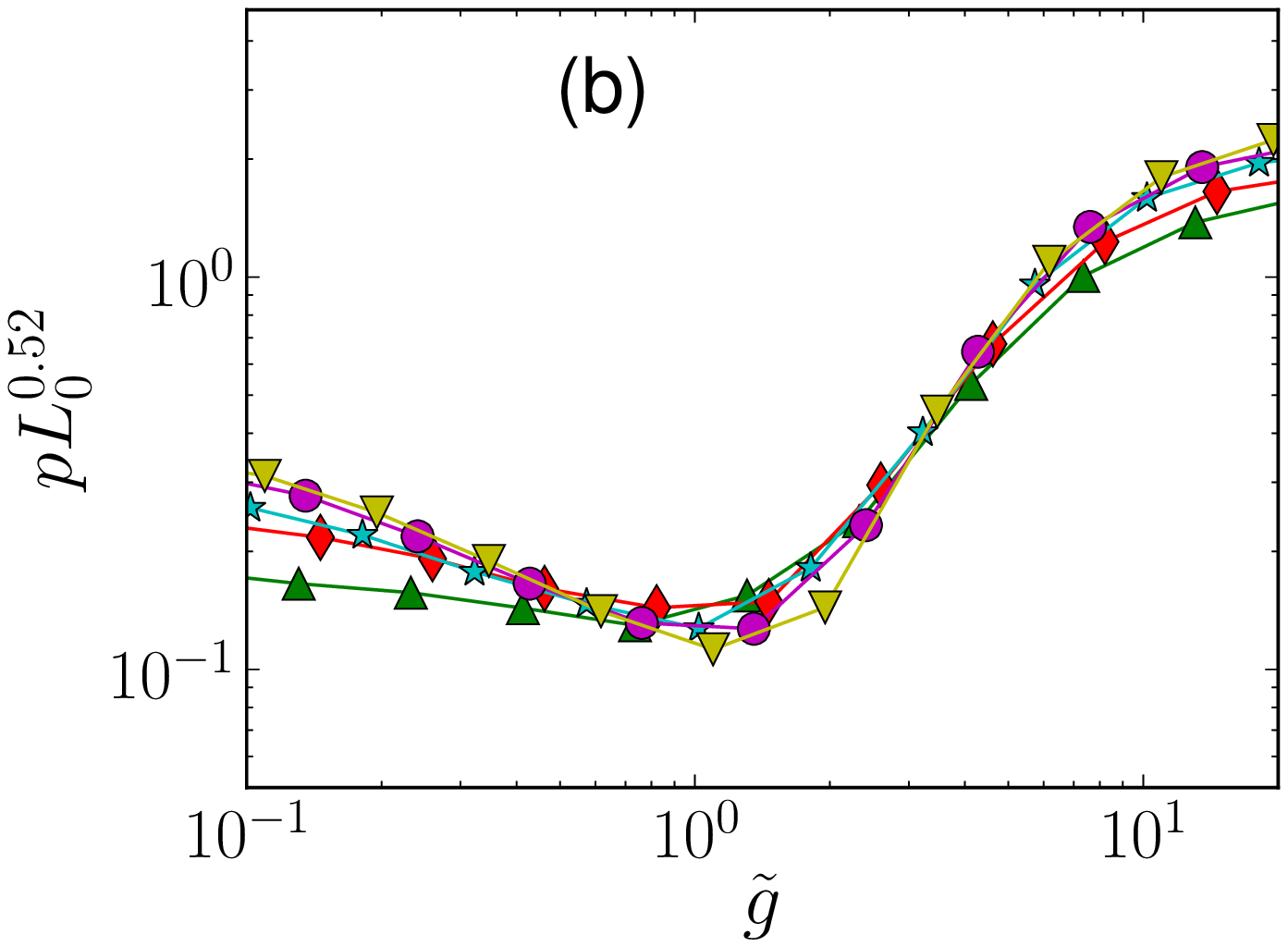}\quad
\includegraphics[width=0.3\columnwidth]{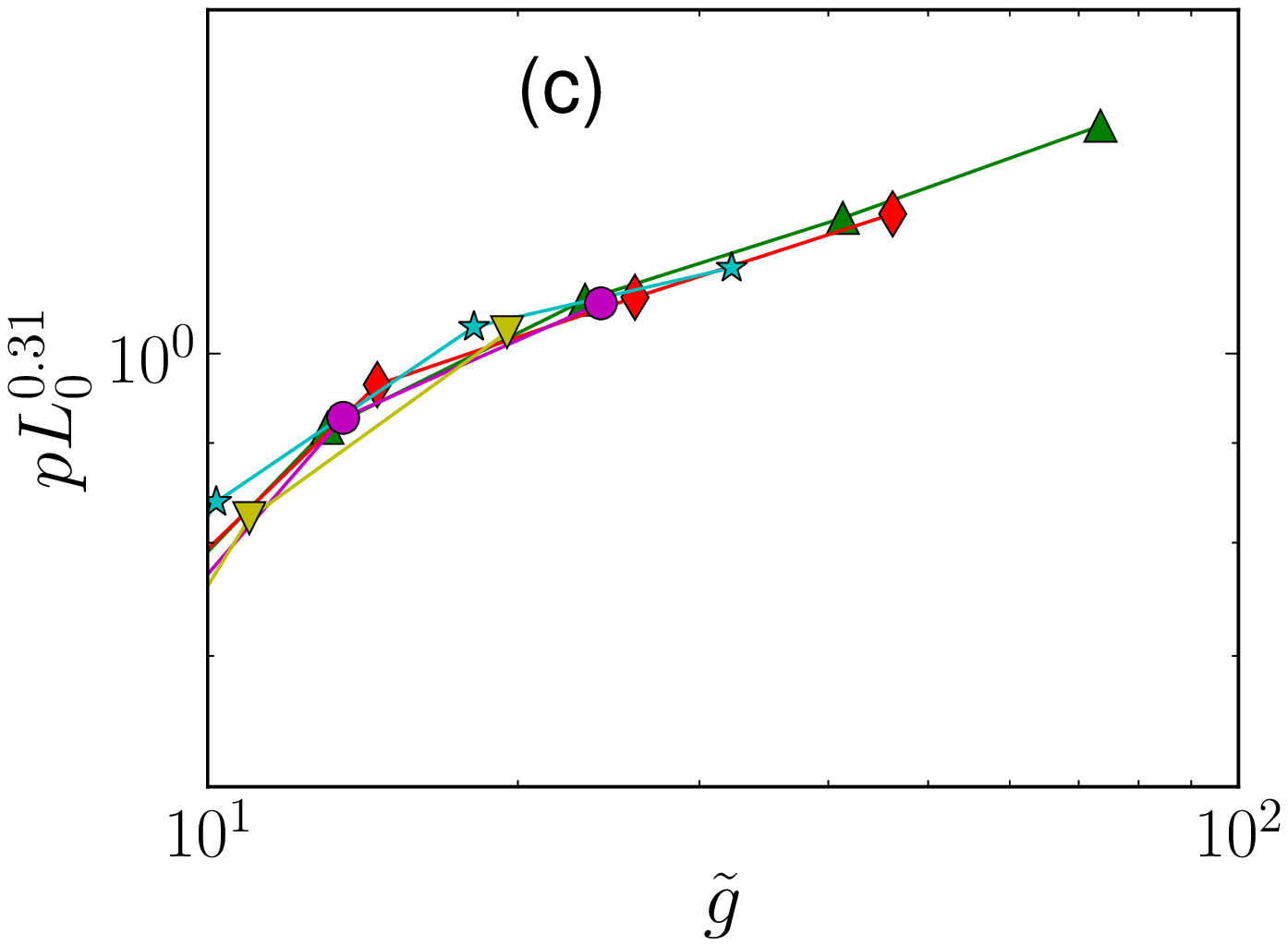}\\
\includegraphics[width=0.3\columnwidth]{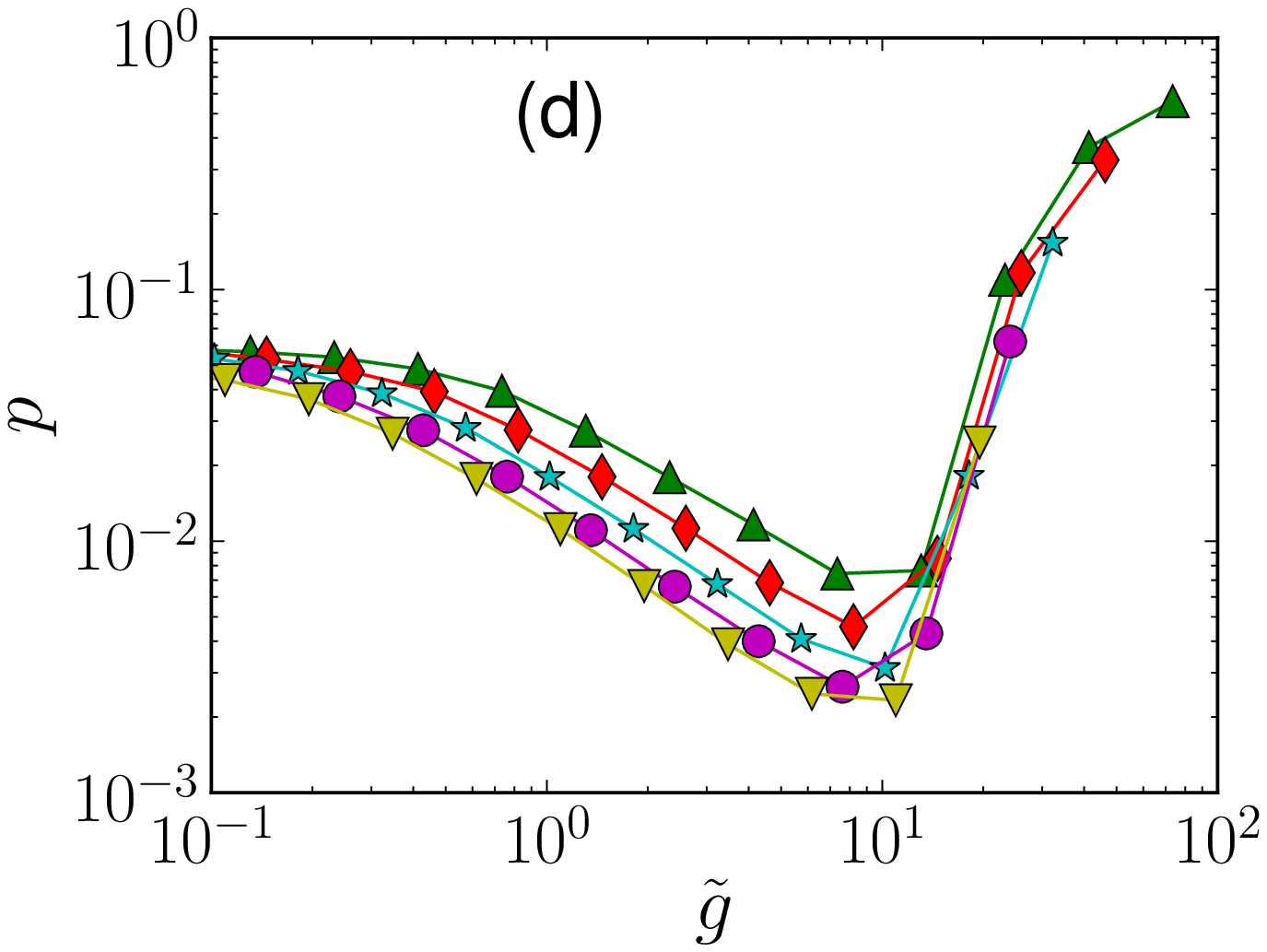}\quad
\includegraphics[width=0.3\columnwidth]{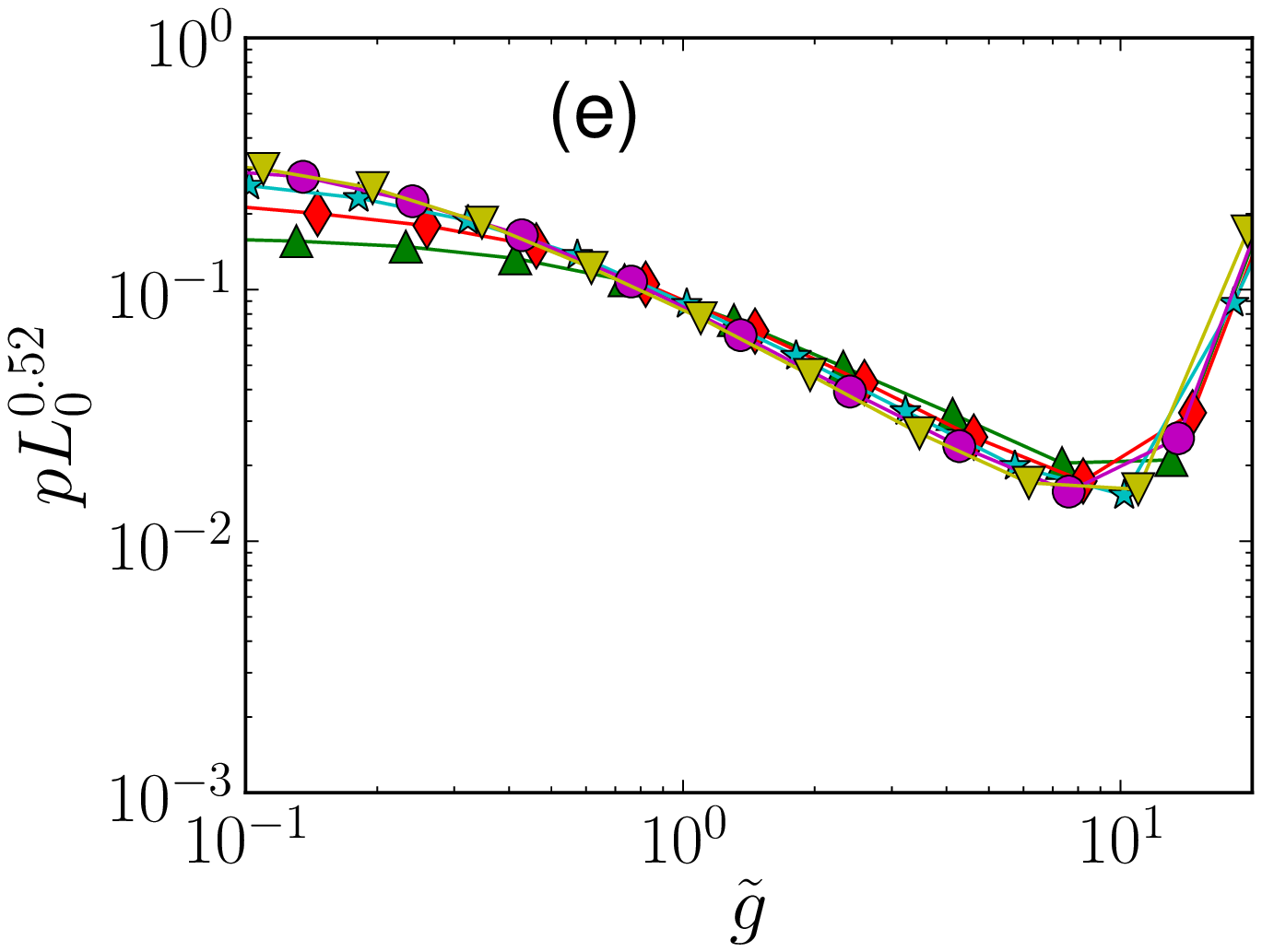}\quad
\includegraphics[width=0.3\columnwidth]{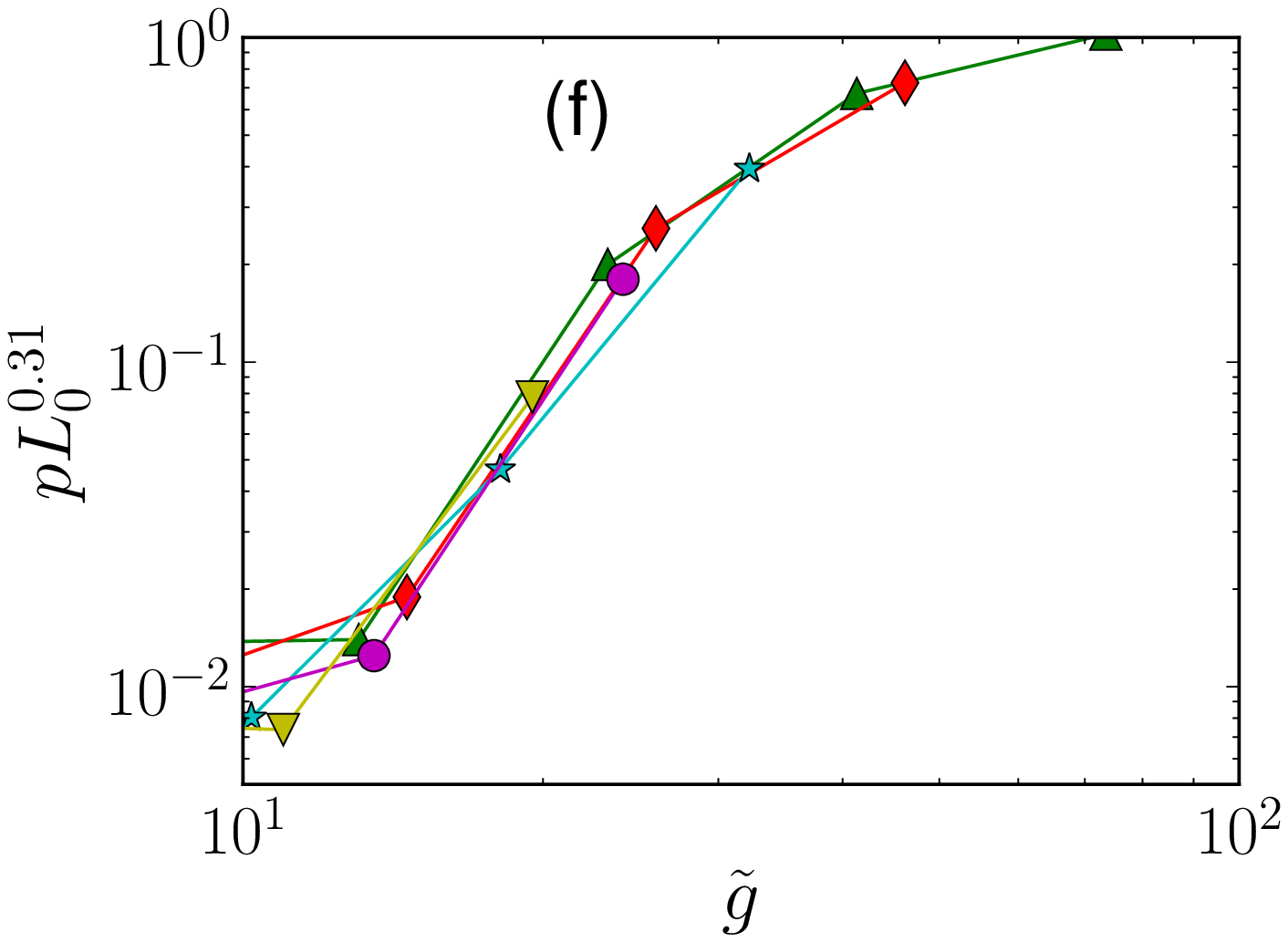}

\caption{\label{fig:pvsgt}(Color online) Scaling of the survival probability
$p(t=10^{5})$ (log-log plot) as a function of the scaled nonlinearity
$\tilde{g}$. (a-c) $\gamma=10^{-5}$ and $W=4$, (d-f) $\gamma=10^{-3}$,
$W=1$. Left column: $p$ not scaled and full range in $\tilde{g}$;
middle column: $p$ scaled in $L_{0}^{0.52}$ zoomed on the chaotic
region $0.1\le\tilde{g}\le10$; right column: $p$ scaled in $L_{0}^{0.31}$
zoomed on the self-trapping region $\tilde{g}>10$. Values of $L_{0}$
: 7 (green triangles), 13 (red diamonds), 21 (cyan stars), 31 (magenta
circles), 41 (yellow inverted triangles).}
\end{figure}

Figure~\ref{fig:pvsgt} shows that the probability $p(t)$ can
also be approximately scaled in $L_{0}$, but the scaling depends
on the dynamical regime: $\tilde{p}=pL_{0}^{\nu}$ with $\nu=0.52$
in the chaotic regime and $\nu=0.31$ in the self-trapped regime (and,
trivially, $\nu=0$ in the quasilocalised regime).

\section{Conclusion}

We have thus characterised the influence of decoherence on the dynamics
of interacting bosons in a disordered potential. Spontaneous emission
changes considerably the time behaviour of the survival probability,
but the use of the scaled parameter $\tilde{g}$ takes into account
the crucial sensitivity of the system to the initial conditions, even
in presence of decoherence induced by spontaneous emission, allowing
a global characterisation of the dynamics. This shows that the effects
of interactions and decoherence tend to add up rather than compete.
These above results put into evidence the usefullness of the scaling
on the initial state as a tool allowing state independent characterisation
of the dynamics, which thus constitutes an important element of a {}``new language'' better adapted to describe the quantum mechanics
of nonlinear systems.

Laboratoire de Physique des Lasers, Atomes et Mol\'ecules is Unit\'e
Mixte de Recherche 8523 of CNRS. Work partially supported by Agence
Nationale de la Recherche's LAKRIDI grant and {}``Labex'' CEMPI.

%\bibliographystyle{unsrt}
%\bibliography{ArtDataBase}

\begin{thebibliography}{10}

\bibitem{Anderson:LocAnderson:PR58}
P.~W. Anderson.
\newblock {Absence of Diffusion in Certain Random Lattices}.
\newblock {\em Phys. Rev.}, 109(5):1492--1505, 1958.

\bibitem{Billy:AndersonBEC1D:N08}
J.~Billy, V.~Josse, Z.~Zuo, A.~Bernard, B.~Hambrecht, P.~Lugan, D.~Cl{\'e}ment,
  L.~Sanchez-Palencia, P.~Bouyer, and A.~Aspect.
\newblock {Direct observation of Anderson localization of matter-waves in a
  controlled disorder}.
\newblock {\em Nature (London)}, 453:891--894, 2008.

\bibitem{Inguscio:AubryAndreInteractions:NP10}
B.~Deissler, M.~Zaccanti, G.~Roati, C.~d'Errico, M.~Fattori, M.~Modugno,
  G.~Modugno, and M.~Inguscio.
\newblock {Delocalization of a disordered bosonic system by repulsive
  interactions}.
\newblock {\em Nat. Phys.}, 6:354--358, 2010.

\bibitem{Kondov:ThreeDimensionalAnderson:S11}
S.~S. Kondov, W.~R. McGehee, J.~J. Zirbel, and B.~DeMarco.
\newblock {Three-Dimensional Anderson Localization of Ultracold Matter}.
\newblock {\em Science}, 334(6052):66 --68, 2011.

\bibitem{Jendrzejewski:AndersonLoc3D:NP12}
F.~Jendrzejewski, A.~Bernard, K.~Muller, P.~Cheinet, V.~Josse, M.~Piraud,
  L.~Pezze, L.~Sanchez-Palencia, A.~Aspect, and P.~Bouyer.
\newblock {Three-dimensional localization of ultracold atoms in an optical
  disordered potential}.
\newblock {\em Nat. Phys.}, 8(5):398--403, 2012.

\bibitem{Moore:AtomOpticsRealizationQKR:PRL95}
F.~L. Moore, J.~C. Robinson, C.~F. Bharucha, B.~Sundaram, and M.~G. Raizen.
\newblock {Atom Optics Realization of the Quantum {$\delta$}-Kicked Rotor}.
\newblock {\em Phys. Rev. Lett.}, 75(25):4598--4601, 1995.

\bibitem{Lignier:Reversibility:PRL05}
H.~Lignier, J.~Chab{\'e}, D.~Delande, J.~C. Garreau, and P.~Szriftgiser.
\newblock {Reversible Destruction of Dynamical Localization}.
\newblock {\em Phys. Rev. Lett.}, 95(23):234101, 2005.

\bibitem{Chabe:PetitPic:PRL06}
J.~Chab{\'e}, H.~Lignier, H.~Cavalcante, D.~Delande, P.~Szriftgiser, and J.~C.
  Garreau.
\newblock {Quantum Scaling Laws in the Onset of Dynamical Delocalization}.
\newblock {\em Phys. Rev. Lett.}, 97(26):264101, 2006.

\bibitem{Casati:IncommFreqsQKR:PRL89}
G.~Casati, I.~Guarneri, and D.~L. Shepelyansky.
\newblock {Anderson transition in a one-dimensional system with three
  incommensurable frequencies}.
\newblock {\em Phys. Rev. Lett.}, 62(4):345--348, 1989.

\bibitem{Chabe:Anderson:PRL08}
J.~Chab{\'e}, G.~Lemari{\'e}, B.~Gr{\'e}maud, D.~Delande, P.~Szriftgiser, and
  J.~C. Garreau.
\newblock {Experimental Observation of the Anderson Metal-Insulator Transition
  with Atomic Matter Waves}.
\newblock {\em Phys. Rev. Lett.}, 101(25):255702, 2008.

\bibitem{Lemarie:CriticalStateAndersonTransition:PRL10}
G.~Lemari{\'e}, H.~Lignier, D.~Delande, P.~Szriftgiser, and J.~C. Garreau.
\newblock {Critical State of the Anderson Transition: Between a Metal and an
  Insulator}.
\newblock {\em Phys. Rev. Lett.}, 105(9):090601, 2010.

\bibitem{Lopez:ExperimentalTestOfUniversality:PRL12}
M.~Lopez, J.-F. Cl{\'e}ment, P.~Szriftgiser, J.~C. Garreau, and D.~Delande.
\newblock {Experimental Test of Universality of the Anderson Transition}.
\newblock {\em Phys. Rev. Lett.}, 108(9):095701, 2012.

\bibitem{Lemarie:AndersonLong:PRA09}
G.~Lemari{\'e}, J.~Chab{\'e}, P.~Szriftgiser, J.~C. Garreau, B.~Gr{\'e}maud,
  and D.~Delande.
\newblock {Observation of the Anderson metal-insulator transition with atomic
  matter waves: Theory and experiment}.
\newblock {\em Phys. Rev. A}, 80(4):043626, 2009.

\bibitem{Chin:FeshbachResonances:RMP10}
C.~Chin, R.~Grimm, P.~Julienne, and E.~Tiesinga.
\newblock {Feshbach resonances in ultracold gases}.
\newblock {\em Rev. Mod. Phys.}, 82(2):1225--1286, 2010.

\bibitem{Vermersch:AndersonInt:PRE12}
B.~Vermersch and J.~C. Garreau.
\newblock {Interacting ultracold bosons in disordered lattices: Sensitivity of
  the dynamics to the initial state}.
\newblock {\em Phys. Rev. E}, 85(4):046213, 2012.

\bibitem{Shepelyansky:DisorderNonlin:PRL08}
A.~S. Pikovsky and D.~L. Shepelyansky.
\newblock {Destruction of Anderson Localization by a Weak Nonlinearity}.
\newblock {\em Phys. Rev. Lett.}, 100(9):094101, 2008.

\bibitem{Flach:DisorderNonlineChaos:EPL10}
T.~V. Laptyeva, J.~D. Bodyfelt, D.~O. Krimer, C.~Skokos, and S.~Flach.
\newblock {The crossover from strong to weak chaos for nonlinear waves in
  disordered systems}.
\newblock {\em EPL (Europhysics Letters)}, 91(3):30001, 2010.

\bibitem{Cohen:LocDynTheo:PRA91}
D.~Cohen.
\newblock {Quantum chaos, dynamical correlations, and the effect of noise on
  localization}.
\newblock {\em Phys. Rev. A}, 44:2292--2313, 1991.

\bibitem{Lepers:SupprSpontEm:PRA10}
M.~Lepers, V.~Zehnl{\'e}, and J.~C. Garreau.
\newblock {Suppression of decoherence-induced diffusion in the quantum kicked
  rotor}.
\newblock {\em Phys. Rev. A}, 81(6):062132, 2010.

\bibitem{Luck:SystDesord:92}
J.~M. Luck.
\newblock {\em {Syst{\`e}mes d{\'e}sordonn{\'e}s unidimensionnels}}.
\newblock {Aléa Sacaly}, {Gif sur Yvette, France}, 1992.

\end{thebibliography}

\end{document}